\documentclass{aa}  

\usepackage{graphicx}
\usepackage{txfonts}
\usepackage{xcolor}

\usepackage{multirow}

\begin{document} 

\title{Important stellar perturbers found during the StePPeD database update based on \textit{Gaia} EDR3 data}

\titlerunning{Stellar perturbers found during the StePPeD database update.}

\authorrunning{P. A. Dybczyński et al.}

\author{Piotr A. Dybczyński
          \thanks{\email{dybol@amu.edu.pl}}
          \and
        Filip Berski
          \and
        Jakub Tokarek 
          \and
        Edyta Podlewska-Gaca
          \and
        Krzysztof Langner
          \and
        Przemysław Bartczak  
        }

   \institute{Astronomical Observatory Institute, Faculty of Physics, Adam Mickiewicz University, Słoneczna 36, PL60-283 Poznań, Poland}

   \date{Received XXXXXX; accepted XXXXX}

 
  \abstract
   {In 2020, the initial version of the Stellar Potential Perturbers Database (StePPeD) was presented with the aim to deliver up-to-date information on the stars and stellar systems that may perturb a long-period comet motion. We used the minimal distance between a star and the Sun as a selecting tool when compiling a list of interesting objects with close encounters with the Solar System, and our selection for that study was based on \textit{Gaia} DR2 data.}
   {When the \textit{Gaia} EDR3 data release was published, it became necessary to update this database. Additionally, we performed Monte Carlo simulations to obtain uncertainties on the parameters of the closest approach to the Sun of each object.}
   {We recalculated the close approach parameters of all stars in the previous StePPeD release, which resulted in removing approximately one-third of the total. Then we searched for new candidates in the whole \textit{Gaia} EDR3 catalogue. We also take into account the duplicity of the found stars and additionally searched for double stars passing near the Sun which had been overlooked in previous papers. We also found the necessary mass estimates for new objects and updated this information for previously selected stars.}
   {After a careful checking of all the collected data, we composed a new list of 155 potential stellar perturbers of the long-period comet motion. We applied a new threshold of 2~pc for the minimum star-Sun distance. This list consists of 146 single stars and nine multiple systems. For each object, we also estimated the uncertainty of the parameters of their closest approach to the Sun. Among these stars, we found a new potential strong past perturber, \object{HD~7977,} and confirmed the plausibility of a similar action on the part of \object{Gliese~710} in the future. }
   {}

\keywords{comets:general -- stars:kinematics and dynamics  -- (Galaxy:) solar neighbourhood}
   \maketitle

\section{Introduction}

Stellar perturbations have been the only mechanism producing observable comet orbits proposed by \citet{oort:1950}. The importance of the galactic perturbations has been recognised since then as the second most important factor, but growing knowledge on the nearby stars positions and movements causes the necessity to simultaneously account for both these sources of perturbations in the studies of long-period comets (hereafter LPCs) motion and origin. \citet{Dyb-Breiter:2021} recently proposed a new and efficient method for such calculations, which is partially based on the data on potential stellar perturbers.

In \citet{rita-pad-magda:2020} (WDP20) the authors introduced a publicly available Stellar Potential Perturbers Database (StePPeD)\footnote{\url{https://pad2.astro.amu.edu.pl/StePPeD}} containing data on potential stellar perturbers of  LPCs motion. This database has had four revisions between April and July 2020, which finally led to release 2.3. However, after the {\it Gaia} Early Data Release 3 was made available \citep{GaiaEDR3-summary:2021}, it appeared that the substantial update of the StePPeD database had become necessary.

In this paper, we describe the update of the StePPeD database, which fully  accounts for the {\it Gaia} EDR3 results \citep{GaiaEDR3-summary:2021}.  Motivated by several important differences between astrometric data in the last two 
Gaia data releases, as well as the large uncertainties for some objects
and the expected significant improvement in the near future, we decided to temporary limit our list of potential perturbers by applying a smaller threshold for the minimal star-Sun  distance that is used to select objects of interest. The previous threshold of 4~pc resulted in a list of 825 objects and now, using a limit of 2~pc, we shortened the list to 155 stars or stellar system.

The update process first concentrated on a new data for the perturbers from the old list and this step is described in Section~\ref{update_part1}. In Section~\ref{sect:update_part2}, we describe the procedure of finding new candidates among stars in {\it Gaia} EDR3 catalogue. It also presents how we deal with double stars and discusses sources of additional data (e.g. stellar mass estimates). Moreover, we discuss various sources of the uncertainties of our results. Section~\ref{final-passages} presents the obtained results for the closest star-Sun encounters. The most important case of the star P0230 (HD~7977) is discussed in Section~\ref{sect:P0230_case}.  The next section answers the question on the influence of such a close passage of a star near the Sun on the Solar System planets and small bodies. Section~\ref{sect:other_perturbers} briefly describes other important stellar perturbers. The final section summarises our work on the StePPeD database update.
   
\section{Update part I: Refinement of the star list included in the StePPeD release 2.3}\label{update_part1}

The StePPeD release 2.3, made available on July 27, 2020, consisted of 825 potential perturbers, that is, 787 single stars and 38 multiple systems. Among them, 648 objects might nominally pass the Sun closer than 4~pc. As it concerns multiple systems, systemic position and velocities were calculated from individual component data for 34 systems. For the remaining four (P1008, P1024, P1037, P1069), the data were taken directly from dedicated papers (see WDP20 for details). Astrometric data were gathered for 858 individual stars, 833 of them were taken from {\it Gaia} DR2 catalogue \citep{Gaia-DR2:2018}, and for remaining 25 from other sources.

The first step in updating StePPeD database was to search for all these stars in the {\it Gaia} EDR3 catalogue. It was rather troublesome since in many cases cross-identifications were ambiguous -- in many cases, we found two stars in EDR3 possibly corresponding to one star in DR2. It resulted often from finding a possible second component of a double system that was absent or overlooked in previous investigations. Finally, we identified 43 new pairs as suspected to be binaries, many of them with poor quality or incomplete data (missing mass estimate or radial velocity).

The comparison of data between DR2 and EDR3  led to surprising results. We were able to compare 836 stars  (single perturbers and components of the stellar systems)  from our potential perturbers list (which are present in both of these catalogues). Among these, we found 197 stars with a parallax difference greater than 100\% (and additionally nine stars with a negative parallax in EDR3). For proper motions, the situation is only slightly better: 148 stars with the proper motion difference in the right ascension  greater than 100\% and 154 stars for declination. Among these 836 stars, 350 have a renormalised unit weight error (RUWE) parameter greater than 1.4, which indicates an unresolved binarity or potentially less accurate data \citep{Gaia-EDR3-validation}. For 43 stars on our list, there are no parallax neither proper motion in EDR3.

After a careful verification of our list, we ended up with 743 single stars and 81 multiple systems. In the new StePPeD release 3.0, due to poor quality or incomplete data, we decided to omit 16 single stars and 20 multiple systems, usually because EDR3 reports much smaller parallax or significantly changed proper motions with respect to the previous data or because they have possible secondaries.  Additionally, for the same reasons, for 11 systems, we decided to use the data for only one component. Over 50 individual comments on particular objects are available at the database web interface. All omitted stars or components were kept in our database for an upcoming verification when the new data become available.

Apart from the astrometry, we also collected radial velocities and mass estimates. The velocities are mainly incorporated from {\it Gaia} DR2, but also from \cite{GaiaDR2-velocity-standards}, along with tens of other individual papers. The mass estimates are most often taken from \cite{Anders:2019} and \citep{TIC-8:2019}, as well as other relevant papers. It is worth mentioning that every numeric input data included in the StePPeD database can be directly inspected through the web interface where  its source is also explicitly shown. There is also a possibility to download all input data as the plain ASCII text file. Based on the acquired data, we calculated the Galactic rectangular heliocentric position and velocity components for 783 perturbers (single stars or the centre of mass of multiples). For an additional five objects (P0403, P1008, P1024, P1037, and P1069) we used systemic data taken from individual papers. All the positions and velocities were moved to the common epoch of 2016.0 -- which is that of the {\it Gaia} EDR3 catalogue.

The above described data for 788 potential LPCs motion perturbers were used as the starting data for backward or forward numerical integration, taking into account all mutual interactions and the overall Galactic potential. For details of our dynamical model, used constants, reference frame, and Galactic potential approximation, we refer to \cite{dyb-berski:2015}. 

Based on the nominal parameters of close encounters with the Sun in WDP20, the authors selected 642 objects which, in the past or future, could approach closer than 4~pc from the Sun. After some revisions, in the StePPeD release 2.3 there are 648 such perturbers. The significant change between {\it Gaia} DR2 and EDR3 in astrometric data for large number of stars from our list resulted in remarkable downsizing of this number. Using the same threshold of 4~pc, we obtained only 407 objects. As many as 157 objects have their nominal closest distance from the Sun greater than 100~pc and among them, 25 have passed or will pass farther than 1~kpc. Some of them, especially the multiple systems, require further verification when new data become available, but most of them will be probably permanently removed from the StePPeD future releases.

A similar work on updating the StePPeD version 2.3 parameters of the closest star-Sun approaches was published recently by \cite{Bobylev-Bajkova:2021} but it is limited to stellar passages closer that 1~pc. They also ignore the multiplicity of the passing stars and use different Galactic parameters and potential model, so they obtained slightly different results.
In November 2021, the radial velocities of a dozen stars have been updated and as a result StePPeD version 3.1 was released (see footnote 1). For details see the 'Changelog' available at the StePPeD Web page.

\section{Update part II: Adding new stars based on \textit{Gaia} EDR3}
\label{sect:update_part2}
Taking into account a large percentage of significant differences in parallaxes and proper motions between {\it Gaia} DR2 and EDR3, as well as the resulting number of perturbers removed from our list, we expect that some new candidates can be found in {\it Gaia} EDR3. To search for new stars, we decided to repeat the whole procedure of finding stars approaching the Sun, similar to what was described in WDP20, in addition to the stars that are already in our list.

Using a linear approximation and the ADQL query similar to that proposed in \cite{Bailer-Jones:2018}, we searched the whole {\it Gaia} EDR3 catalogue \citep{GaiaEDR3-summary:2021} for the stars that have visited or will visit the solar neighbourhood. However, this time we did not restrict our calculations to objects with a nominal linear minimal distance from the Sun $D_{min}^{lin}<10$~pc, as proposed in some previous attempts \citep{dyb-berski:2015,Bailer-Jones:2018}, but we take into account also stars that approach the Sun (according to a linear model) up to a heliocentric distance of 280~pc. The reason we would look for objects with $D_{min}^{lin}>10$ pc is that the track of the Sun and stars in longer time intervals are significantly different from a straight line and linear approximation cannot provide reliable results. Unfortunately, the longer time of motion of objects increases the error of our results  due to the propagation of the stellar data uncertainties. That issue and the low number of remaining objects after numerical verification (only 14 stars with $D_{min}^{lin}>200$~pc and $D_{min}^{lin}\leq280$~pc) prompted us to stop searching for stellar perturbers with $D_{min}^{lin}$ values larger than 280~pc.

\begin{figure}
        \includegraphics[width=0.35 \textheight]{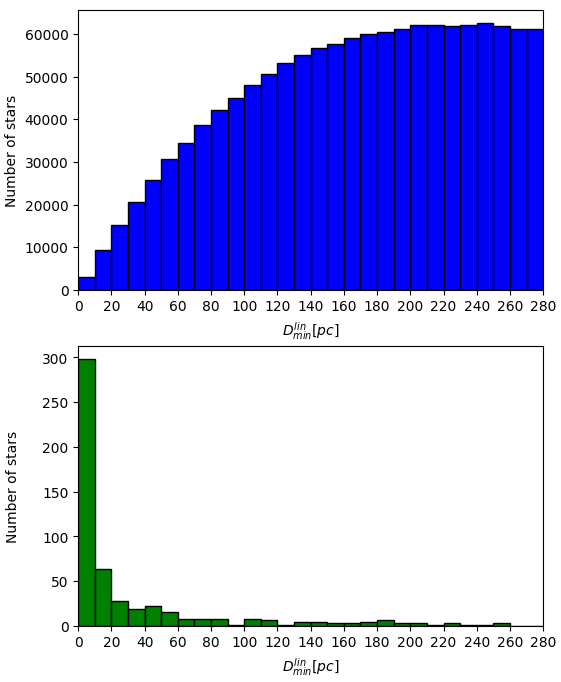}
        \caption{New stellar candidates statistics. Top histogram shows a number of new candidates for close flyby stars from {\it Gaia} EDR3 catalogue based on a linear motion approximation. The bottom panel shows a number of stars remaining after numerical integration and applying a threshold of for the nominal star distance $D_{min}^{nom} \leq 5$~pc. The horizontal axis describes a minimal distance between the star and the Sun $D_{min}^{lin}$ calculated from a linear approximation.}
        \label{fig-number-of-stars-dl}
\end{figure}

Using a linear approximation we find over 1.3 million of objects with $D_{min}^{lin} \leq 280$~pc. Among them we find over three thousands stars with $D_{min}^{lin} \leq 10$ pc~and over nine thousands stars with $10~pc < D_{min}^{lin} \leq 20$~pc. The number of potential candidates grows significantly with the allowed $D_{min}^{lin}$. At $D_{min}^{lin} = 210$~pc this growing trend almost stops and a number of stars per 10~pc bin stabilises around 60 thousands of objects, with a maximum value of 62521 stars for $240~pc < D_{min}^{lin} \leq 250$~pc. The details are shown in top panel of Fig. \ref{fig-number-of-stars-dl}.

The next step is to numerically integrate the stellar motion and obtain much more reliable proximity distance value. Due to a large number of objects obtained from a linear approximation (over 1.3 million stars), we decided  to first use a simple model. It ignores any mutual action between a star and the Sun, therefore, their motion depends only on the gravitational potential of the Galaxy. To compensate the simplicity of this approach we decided to slightly increase the threshold for star-Sun minimal distance of potentially interesting objects from 4~pc used in the past to 5~pc. Our investigations show that the relative error between this simple model and the model including distance corrections and interactions between bodies is about 13.5\%. For example, for star \object{EDR3 469547750615777280, the} minimum distance changes from 4.61~pc to 3.07~pc when we use a more accurate model.

Unexpectedly, the numerical integration showed that many of objects, that pass much further than 10~pc from the Sun according to the linear approximation, can in fact approach the Sun at a very close distance. Please note that we discuss here only the nominal results (see the bottom panel of Fig. \ref{fig-number-of-stars-dl}).

For example, the star \object{EDR3~4312257326836128896} (P2001), which in a linear approximation has the minimal distance $D_{min}^{lin} = 53.12$~pc appeared to pass the Sun at $D_{min}^{nom} = 0.38$~pc based on a numerical integration. The star \object{EDR3~6755981602406174720} (P2009) is even a more spectacular example: we obtained a change from $D_{min}^{lin} = 252.28$~pc to $D_{min}^{nom} =  0.83$~pc. We also recorded several examples of the opposite effect: for more that one hundred of stars a linear approximation gives the minimal distance D$_{min}^{lin} < 10$~pc, while the nominal distance from the numerical integration is D$_{min}^{nom} > 100$~pc. The extreme example is the star \object{EDR3~450118246277153152} for which D$_{min}^{lin} =7.9$~pc, but D$_{min}^{nom} = 4541$~pc.

Following this step, we obtained 527  new star flybys near the Sun closer than 5~pc (all stars or stellar systems present on our previous list were automatically omitted).  After we performed the numerical integration, it appeared that only about 10\%\ of objects with $D_{min}^{lin} \leq 10$~pc have $D_{min}^{nom}$ smaller than 5~pc; namely, it was 298 from 3121 tested stars. In the next bin ($10~pc < D_{min}^{lin} \leq 20$~pc), we found 64 objects from among over nine thousands of candidates.  Finally, as many as a half of all new perturbers have $D_{min}^{lin} > 10$~pc and 90\%\ of all objects  have $D_{min}^{lin} \leq 100$~pc. The largest value of $D_{min}^{lin}$, which was reduced below the 5~pc threshold as a result of a numerical integration is for the star \object{EDR3~5616156518241119872}. In this case, we have $D_{min}^{lin} = 258.68$ pc and $D_{min}^{nom} = 4.01$~pc. 

\subsection{Searching for secondaries}

At least 60 of our targets appeared to be  members of binary systems listed in the Gaia EDR3 
catalogue of possible binary stars of \cite{million-binaries}.
Furthermore, by inspecting carefully the vicinity of each star in 
our list we found some additional possible binaries.
For example, the star  \object{EDR3~196124257030748544} (P2024) have probably a companion \object{EDR3~196124257030748032} at the angular distance of 13.7~arcsec. Thanks to the fact that the secondary is relatively bright ($\sim$14 mag), we were able to collect a complete set of data (masses, radial velocities) for both components. However, the physical binarity of this system is highly uncertain.

In the case of a binary system, the knowledge of all parameters is crucial, because it is necessary to calculate the spatial position and velocity of its centre of mass. In such cases, the movement of a system as a whole might be completely different from a single component motion in space. It is worth mentioning that most of the binaries listed in \citep{million-binaries} are faint objects with a brightness fainter than 18 mag, so a complete set of parameters is difficult to find. In particular, lacking are radial velocities or mass estimates.

\subsection{Unrecognised binaries passing close to the Sun}
\label{binaries}

In general, there is also the possibility that we have completely overlooked a binary passing close to the Sun when both components (treated separately) have their minimal distance greater than our adopted threshold, but the mass centre of such a system could approach the Sun closely. To our knowledge it is the first attempt to find such systems. 

\begin{figure}
        \includegraphics[width=0.9 \columnwidth]{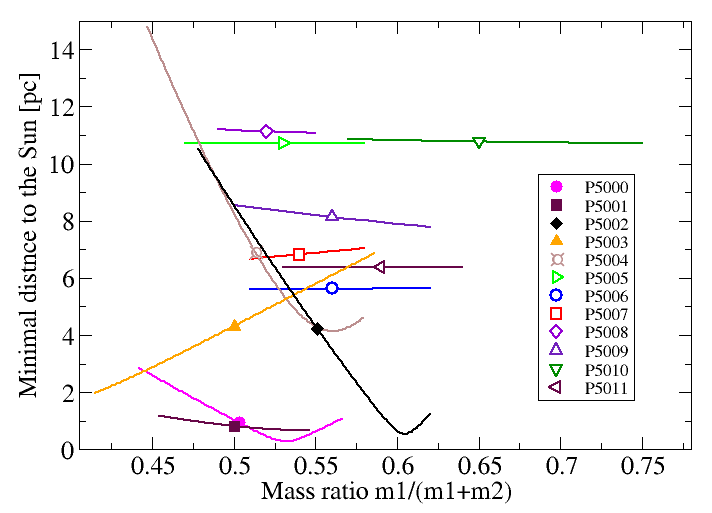}
        \caption{Minimum distance from the centre of mass of binary systems to the Sun, including the possible error intervals arising from the uncertainties in the mass estimates of the individual components (see Table~\ref{tab-binaries-15})}
        \label{double_near15}
\end{figure}

We searched for candidates in \citep{million-binaries}. Our first  requirement is that radial velocities for both components are known. In applying this criterion, we found 15 359 appropriate candidate binaries.  For all systems on this shortened list, we calculated their centres of masses and velocities based on the dependence of the mass distribution varying from 0 to 1 with step 0.001, where: 1 means all the mass of the system is located in component A and 0 is that all the  mass of the system is located in component B. At this point, the real masses of components are not necessary, we only need a mass ratio to calculate the centres of mass position and velocity. That allows us to determine their trajectories in the Galaxy. 
We found only 156 systems of interest. For them we searched for real mass estimates of components. A list of the used mass sources is described in Section \ref{num-integ-2}. We found mass estimate for both components for only 108 systems, and for the remaining 48 systems, we found no mass or mass for only one component. Using the real mass estimates we found that from the binaries with complete data only 12 systems might pass closer than 15 pc to the Sun, within their mass uncertainties (see Figure \ref{double_near15}). We used here a larger threshold of 15~pc because close approach parameters strongly depend on the components' masses, which often come with large uncertainties. Using the nominal dynamical parameters for both components, we calculated how the uncertainty in mass estimates impacts the minimal system distance from the Sun. These systems parameters are presented in Table \ref{tab-binaries-15}. Consecutive columns content is as follows: ID in the StePPeD database; first component {\it Gaia} EDR3 identification number; its mass with the 1$\sigma$ uncertainty; next two columns contained the same information for the second component. In the last two columns we present our results as the system minimum close-up distance and the time of that event. The positive time indicates that the approach will happen in the future, the negative describes the approaches in the past. The errors of last two parameters are here calculated only from the uncertainties in mass estimations.

For all cases, we have studied the effect of the ratio of the masses of the system components on the distance for transits through the Solar System. We selected three potential candidate systems for which only one mass is available. In these three cases we found the optimal mass distribution between components, which allows them for a Solar System pass-by below 15~pc (pairs: [\object{Gaia~EDR3 2738129612130247680}, 2738129616426300160], [\object{Gaia EDR3~3154610636614026624}, 3154610636617646848], and [\object{Gaia EDR3 3228938722162512256}, 3228938722164251904]). This candidates will be valid targets for further calculations of the approaches to the Sun if in future both mass estimations will become available.

\begin{figure}
        \includegraphics[width=0.35 \textheight]{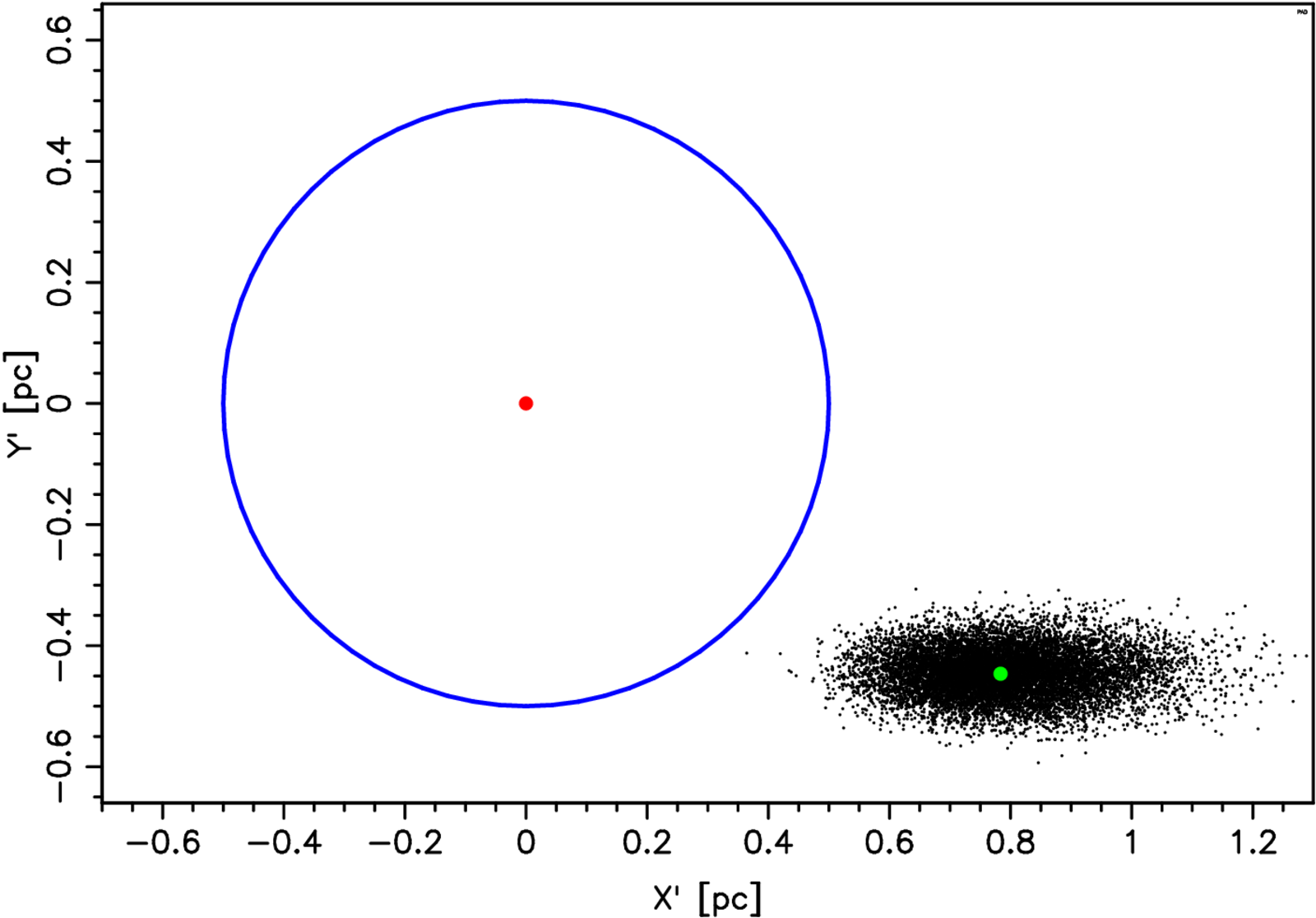}
        \caption{Picture represents projection of 11 635 VSs of P5000 system (black dots) in their closest encounter with the Sun (red dot). Nominal position of P5000 is marked with green dot. Blue circle is showing outer Oort Cloud boundary at 0.5 pc away from the Sun. All VSs always have the same masses, but the dynamical parameters have been randomised.}
        \label{P5000clones}
\end{figure}

In Figure \ref{P5000clones}, we show an example case of P5000 (\object{BD-06~310}, \object{WDS~J01395-0612}). In this system, both components have mass estimates in the TESS-8 catalogue \citep{stassun2018tess}. For their nominal masses, we created 11 635 random virtual stars (hereafter VSs) with different dynamical parameters randomly drawn according to the covariance matrix from {\it Gaia} EDR3. Our calculations show that in 6.8 million years, P5000 will fly past the Sun at a distance of 0.9 pc. Statistics of that and other closest binary systems flybys are presented in Table \ref{tab-binaries-closest}. Due to the uncertainty in mass estimates, this cloud of VSs could shift towards the Sun or in an opposite direction in the range shown in the sixth column of Table \ref{tab-binaries-15}.

\begin{table*}
        \caption{Closest approaches of new binary systems. The first column shows the ID number in the StePPeD database, the second {\it Gaia} EDR3 identifiers for both components, the third presents the distance D$^{geom}_{min}$ from the Sun to the centroid of  VSs  with its uncertainty (see text for detail description, Section \ref{stellar-uncert}), the next three columns present a statistical description of the minimal distance, D$^{stat}_{min}$, time of the approach T$^{stat}_{min}$ and the relative velocity, V$^{stat}_{rel}$, for the set of all VSs. In the last column the mass estimate is included.  Here and elsewhere, some extra significant digits are given for uniformity.} \label{tab-binaries-closest}
        
        \begingroup
    \setlength{\tabcolsep}{6pt} 
    \renewcommand{\arraystretch}{1.7} 
    \centering
        \begin{tabular}{c l c  r c c c r} 
                \hline 
                StePPeD & \multicolumn{1}{c}{{\it Gaia} EDR3}   & \multicolumn{1}{c}{geometry}         & \multicolumn{3}{c}{statistics}                        & mass estimate \\ 
                ID & \multicolumn{1}{c}{ID}                     &  D$^{geom}_{min}$ [pc]                             &  D$^{stat}_{min}$ [pc] & T$^{stat}_{min}$ [Myr] & V$^{stat}_{rel}$  [km\,s$^{-1}]$        & [M$_{\odot}$] \\
                \hline
                \multirow{2}{*}{P5000}  & 2479549701620909312 & \multirow{2}{*} {0.909$\pm$0.203}  & \multirow{2}{*}{$0.912^{+0.115}_{-0.097}$} & \multirow{2}{*}{$6.841^{+0.311}_{-0.280}$} & \multirow{2}{*}{7.18 $\pm$ 0.29} & 0.65 \\
        & 2479549701621101184 & & & & & 0.64 \\
                \multirow{2}{*}{P5001}  & 4569088021684763904 & \multirow{2}{*}{0.920$\pm$1.791}  & \multirow{2}{*}{$1.199^{+0.754}_{-0.554}$} & \multirow{2}{*}{$8.732^{+0.295}_{-0.279}$} & \multirow{2}{*}{60.32 $\pm$ 1.09} & 0.77 $^a$ \\
        & 4569088021688882048 & & & & & 0.77 \\
                \multirow{2}{*}{P5002}  & 2182863462586498432 & \multirow{2}{*}{4.139$\pm$0.590}  & \multirow{2}{*}{$4.143^{+0.527}_{-0.515}$} & \multirow{2}{*}{$-33.338 \pm 0.039$} & \multirow{2}{*}{9.946 $\pm$ 0.013} & 1.33 \\
        & 2182863492643748096 & & & & & 1.08 \\
                \multirow{2}{*}{P5003}  & 4470126821233210112 & \multirow{2}{*}{4.437$\pm$0.885}  & \multirow{2}{*}{$4.511^{+0.202}_{-0.183}$} & \multirow{2}{*}{$-14.637^{+0.491}_{-0.531}$} & \multirow{2}{*}{17.15 $\pm$ 0.61} & 1.42 \\
        & 4470126821224457344 & & & & & 1.42 \\      
                \hline
        \end{tabular}
        \tablefoot{$^a$ The mass of this component was assumed  equal to the second component. This is a reliable assumption due to the comparable observed magnitudes of both stars.}
        \endgroup
\end{table*}

\begin{table*}
        \caption{Effect of mass uncertainty on the distance and time of approach to the Sun. The first column shows StePPeD ID, the second and fourth {\it Gaia} EDR3 identifiers of the components. The third and fifth columns contain information on masses and their uncertainties for the components. The last two columns show the minimum distance and the time of approach of the centre of mass of the system to the Sun. Uncertainties in the distance and time of approach to the Sun are here only due to the uncertainties in the determination of the mass components of the binary system.} \label{tab-binaries-15}
        \begingroup
    \setlength{\tabcolsep}{4pt} 
    \renewcommand{\arraystretch}{1.7} 
    \centering
        \begin{tabular}{c l c l c c c } 
                \hline 
                StePPeD & \multicolumn{2}{c}{First component}   & \multicolumn{2}{c}{Second component} & \multicolumn{2}{c}{Mass centre approach to the Sun} \\ 
                ID & \multicolumn{1}{c}{{\it Gaia} EDR3 ID} & mass est. [M$_{\odot}$]&  \multicolumn{1}{c}{{\it Gaia} EDR3 ID} & mass est. [M$_{\odot}$]&  min dist. [pc] & min time [Myr]    \\
                
                \hline
                P5000 & 2479549701620909312 & $0.65\pm{0.08}$ & 2479549701621101184 & $0.64\pm{0.08}$ & $0.90_{-0.62}^{+1.95}$ & $6.84_{-0.12}^{+0.09}$ \\
                P5001 & 4569088021684763904 & $0.77\pm{0.08}^a$ & 4569088021688882048 & $0.77\pm{0.08}$ & $0.88_{-0.21}^{+0.29}$ & $8.73_{-0.04}^{+0.04}$ \\
        P5002 & 2182863462586498432 & $1.33\pm{0.22}$& 2182863492643748096 & $1.08\pm{0.13} $& $4.16_{-3.62}^{+6.36}$ &$-33.37_{-1.26}^{+1.12}$ \\
        P5003 & 4470126821233210112 & $1.42\pm{0.25}$& 4470126821224457344 & $1.42\pm{0.25} $& $4.45_{-2.48}^{+2.40}$&$-14.62_{-0.31}^{+0.31}$ \\
        P5004 & 1031783261988885376 & $1.15\pm{0.16}$& 1031783639946006912 & $1.09\pm{0.14} $& $6.95_{-2.81}^{+7.82}$ &$24.74_{-0.91}^{+1.03}$ \\
                P5005 & 6828177528743255552 & $0.73\pm{0.08}$& 6828177528743255424 & $0.66\pm{0.08} $& $10.724_{-0.001}^{+0.002}$ & $3.502_{-0.001}^{+0.004}$\\
                P5006 & 287980516431427200 & $0.80\pm{0.10}$&  287980512135232512 & $0.62\pm{0.08} $& $5.623_{-0.01}^{+0.06}$ &$0.9845_{-0.0001}^{+0.0001}$ \\
        P5007 & 1361208348808624256 &$0.79_{-0.04}^{+0.02}$& 1361208353105063296 & $0.66\pm{0.08} $&$6.82_{-0.15}^{+0.23}$ &$-9.45_{-0.12}^{+0.16}$ \\
        P5008 & 6643765063414984832     & $0.65\pm{0.08}$&  6643765067712454656 & $0.59\pm{0.02} $&$11.12_{-0.05}^{+0.07}$ &$-4.66_{-0.01}^{+0.01}$\\
        P5009 & 6243585681802135808 & $1.17\pm{0.16}$& 6243585681802137728 & $0.91\pm{0.12} $&$8.13_{-0.33}^{+0.42}$ & $-15.18_{-0.05}^{+0.05}$ \\
                P5010 & 2104485016711846656 &$1.23_{-0.34}^{+0.78}$& 2104486489885306752 & $0.66\pm{0.02} $&$10.78_{-0.07}^{+0.08}$&$8.004_{-0.008}^{+0.006}$\\
                P5011 & 1237090738916392704 & $0.96\pm{0.12}$& 1237090738916392832 & $0.67\pm{0.08} $&$6.372_{-0.003}^{+0.005}$&$-0.396_{-0.007}^{+0.005}$\\
        \hline
        \end{tabular}
        \tablefoot{$^a$ The mass of this component was assumed  equal to the second component. This is a reliable assumption due to the comparable observed magnitudes of both stars.}
        \endgroup
\label{table_bin}
\end{table*}

\subsection{Numerical integration:\ Second approximation}
\label{num-integ-2}

The next step towards more reliable nominal data on stellar encounters with the Sun is to perform a numerical integration 
using a more adequate dynamical model. To the overall Galactic potential, we add the mutual interactions between all considered stars and between these stars and the Sun. We also used more sophisticated distance estimations.

This approach make it necessary to have a mass estimate also for the single stars. We updated the mass estimates for all of them, including those from the previous list because a lot of new sources were made available since the publication of WDP20. To this aim we start the search in the recent large data sets: a catalogue TESS-8 \citep{stassun2018tess}, and the stellar parameters derived by the spectroscopic-astrometric-photometric Bayesian STARHORSE software \citep{Anders:2019}, provided in five catalogues (APOGEE DR16, GALAH DR2, GES DR3, LAMOST DR5, and RAVE DR6) \citep{Queiroz2020}. Recently, \citet{Anders22} have used the data from {\it Gaia} EDR3 to improve their previous estimations. Now, their distances are in much better agreement with the results of \citep{B-J-distance:2021} that we have used. 

The majority of stellar distance were adopted from \cite{B-J-distance:2021}, preferring the photo-geometrical estimates.
 For a selected subset of nearby single stars, we used distances from GCNS catalogue \citep{gcns}. In this catalogue, we found distance estimates for 138 stars in main catalogue and for 28 objects in the rejected part of catalogue.

\subsection{Dealing with the stellar uncertainties}
\label{stellar-uncert}

It is very important to estimate the influence of the stellar data uncertainties on the parameters of the close passage near the Sun. In a separate manner, we checked the influence of the mass estimate uncertainty on the binary system results as it is described in Section \ref{binaries}. Here, we describe a procedure used for dealing with astrometric and kinematic uncertainties. To this purpose, we used a star cloning technique similar to that used by \cite{dyb-berski:2015}. Here, the VS is drawn using the respective six-dimensional covariance matrix described in the {\it Gaia} EDR3 documentation\footnote{see the {\it Gaia} EDR3 documentation, paragraph 4.1.7.0.4 on page 196} from which we obtain random values of $\alpha$, $\delta$, $\tilde{\omega}$, $\mu_{\alpha *}$, $\mu_{\delta}$, $\mu_{r}$. There is only one exception: when the uncertainty of the radial velocity is unknown or it is greater then 10\%, we draw the VS parameters from the five-dimensional covariance matrix, keeping the radial velocity equal to the nominal value. To estimate the influence of one star uncertainties on its parameters of the closest passage near the Sun we draw a VS for this star and replace the nominal starting data with the VS parameters. For the rest of the stars, the nominal starting data remain. Then we numerically integrate the whole set of stars under the Galactic gravitational potential, including their mutual interactions and the interaction with the Sun. We repeated this for at least ten thousand of VSs. Such a procedure is extremely time-consuming, but taking into account the mutual gravity of all bodies seems to be necessary since we have observed several close approaches between stars.

We describe the uncertainty in the close approach parameters (distance, epoch, and relative velocity) using a standard deviation of the approximated Gaussian distribution or, in cases when this approximation failed, we use three percentiles: 16\%, the median, and 84\%. For checking whether the parameter distribution might be treated as Gaussian, we used the test of  \cite{Anderson-Darling:1952}.  We also use a geometrical description of the VSs cloud position with respect to the Sun (the details are described in Section~\ref{final-passages}).

\begin{figure}
        \includegraphics[width=1 \columnwidth]{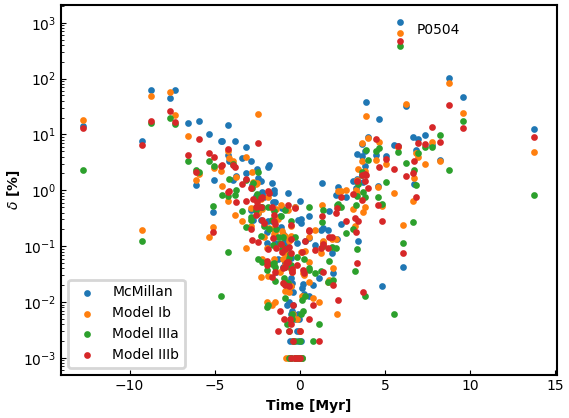}
        \caption{Relative difference in percent of the minimum star-Sun distance between our reference model Ia and four other Galactic potential models described in the text. Horizontal axis describe the epoch of a close encounter. Points in the same colour represent all 155 stars on our list, the colours represent four models of Galactic potential as explained in the picture.}
        \label{models_time}
\end{figure}

\subsection{Stellar close encounters in different Galaxy potential models}
\label{different-models}

Another source of uncertainties is the adopted Galactic potential.
Basing on our previous experience we choose as a reference model the potential described as Ia in \citet{irrgang_et_al:2013} with their numerical parameters. In a recent paper by \cite{Bovy:2021}, this model of a Galactic potential was  presented as the only one which is shown to be in agreement with the Solar System Galactocentric acceleration deduced from the {\it Gaia} EDR3 data \citep{GaiaEDR3-acceleration:2021}. 

To examine the influence of different models of Galactic potential on the minimum distance between stars and the Sun, we integrated all stars back and forth for 50 million years using several different Galactic potential models. The first model is our reference one, referred to as Ia. It is based on the potential proposed by \citet{allen-santillan:1991}, but with the revised parameter values from \citet{irrgang_et_al:2013}. The second model is based on the same formulae, but with parameters taken from \citet[][here Model Ib]{bajkova-bobylev:2017}. Another pair is based on the  Model III from \citet{irrgang_et_al:2013}, which is deduced from the halo density profile of \cite{NFW:1997}.
The former uses parameters
from \citet[][which we call Model IIIa]{irrgang_et_al:2013}, and the latter, from \citet[][here Model IIIb]{bajkova-bobylev:2017}.
Finally, we considered the potential of \cite{mcmillan_2017}, which
is the only non-axisymmetric model in our comparison.

To estimate the effect of using different models of the Galactic potential, we calculate the relative difference with respect to the reference model Ia. Our results show clearly that the difference in minimum distance strongly depends on the time that is necessary for a star to travel from the current position to the closest approach point. It is a combination of effects due to the velocity of the star and its current distance from the Sun. In  Fig.~\ref{models_time} we show the dependence of the relative difference on the proximity distance obtained with four tested models with respect to this time interval.
The largest obtained disparity is for star P0504, the minimum distance obtain from McMillan model is 0.1318~pc; whereas from model Ia, it is only 0.01154~pc, for model Ib is 0.08831~pc, and for models IIIa and IIIb, respectively, it is 0.05533~pc and 0.06531~pc. We see that the maximum difference is about 0.12~pc, while the uncertainty of D$^{nom}_{min}$ resulting from the stellar astrometry errors is about 5.9~pc. We found only one more object that has a relative error bigger than 100 percent, namely P5001 which is a binary star. It is currently 517~pc from the Sun and the close encounter will take place in 8.7 Myr. Almost 90\%\ of stars in our list have a relative error smaller than 10\%. For example, P2002 has a relative difference, for the McMillan model, of 1.38 percent, and for models Ib, IIIa, and IIIb, respectively, it is 0.41, 0.94, and 0.02 percent.

\begin{figure}
        \includegraphics[angle=270,width=1\columnwidth]{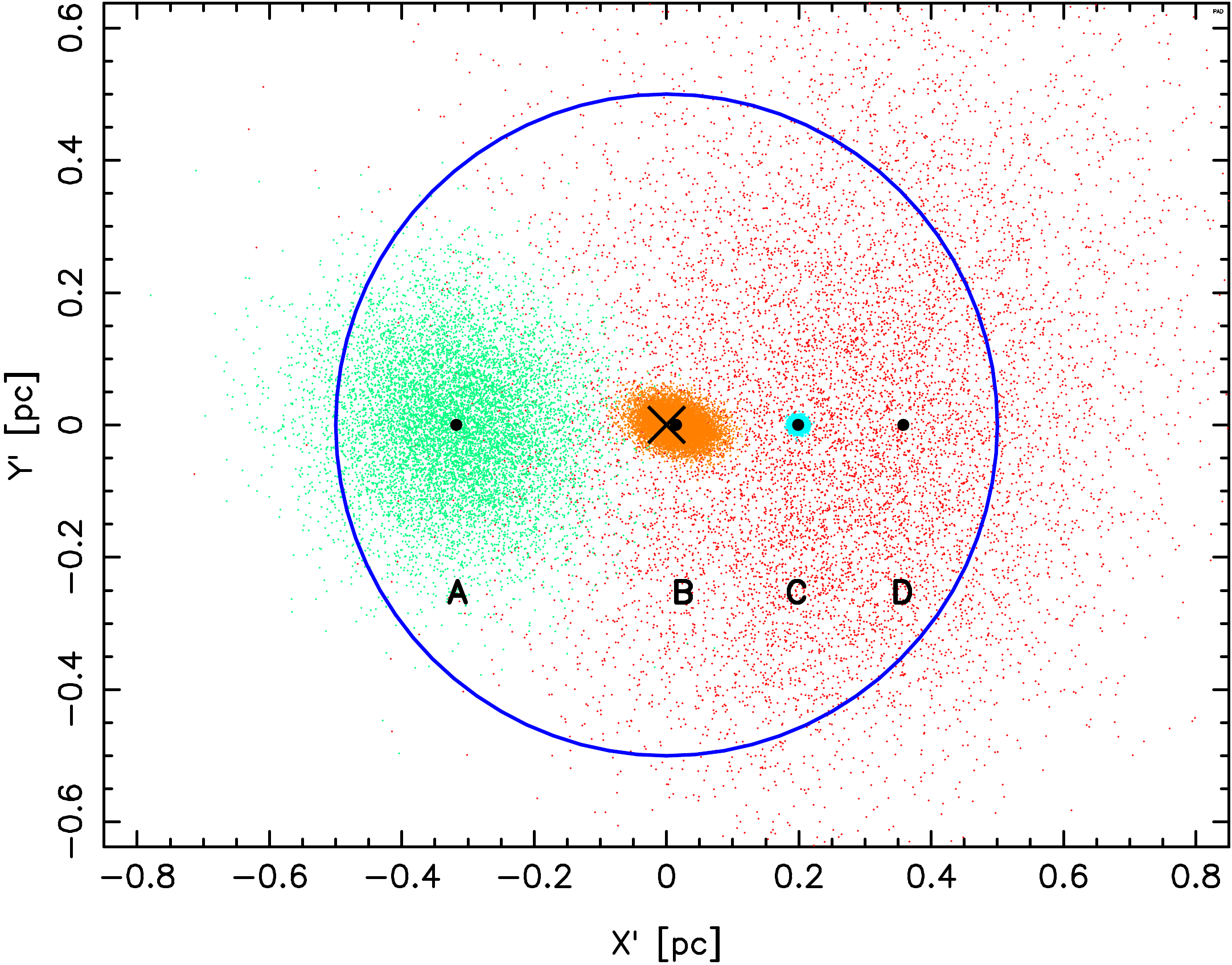} 
        \caption{ First four stars from Table~\ref{tab-closest-past}:\ Comparison of their nominal minimal distances from the Sun and the spread of VSs at their closest approach. Black dots in the centre of each VSs cloud mark the nominal star position at the closest Sun--star approach;  see text on how this composite figure was obtained. Label meanings: A: P0509, B: P0230 (\object{HD~7977}), C:  P0506, D: P0508; see Table~\ref{tab-closest-past} for details on each star.  The blue circle marks the usually adopted outer limit of the Oort cloud and  the cross in the middle marks the Sun's position.}\label{fig-four-stars}
\end{figure}

\begin{figure}
        \includegraphics[width=0.35 \textheight]{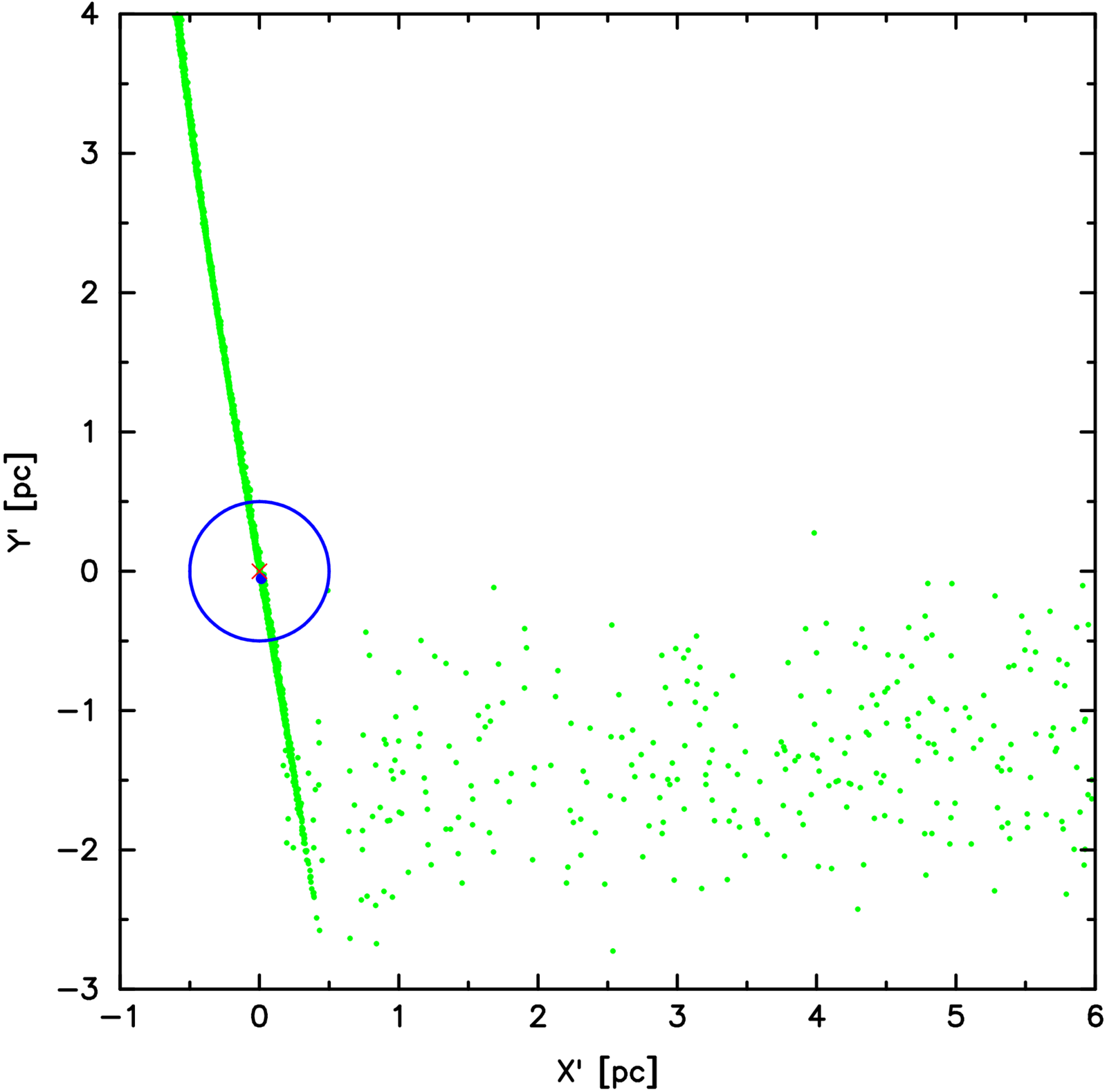}
        \caption{Distribution of P2001 VSs stopped at the closest approach to the Sun and projected on the plane of their maximum scatter. The blue circle represents the boundary of the Oort cloud at 0.5 pc, a red cross is the position of the Sun and the blue dot (very close to the red cross) corresponds to a nominal position of P2001. There are  2655 VSs shown from among 10000 calculated, while the remaining ones are beyond of the image border.}
        \label{P2001clones}
\end{figure}

\begin{figure}
        \includegraphics[width=0.35 \textheight]{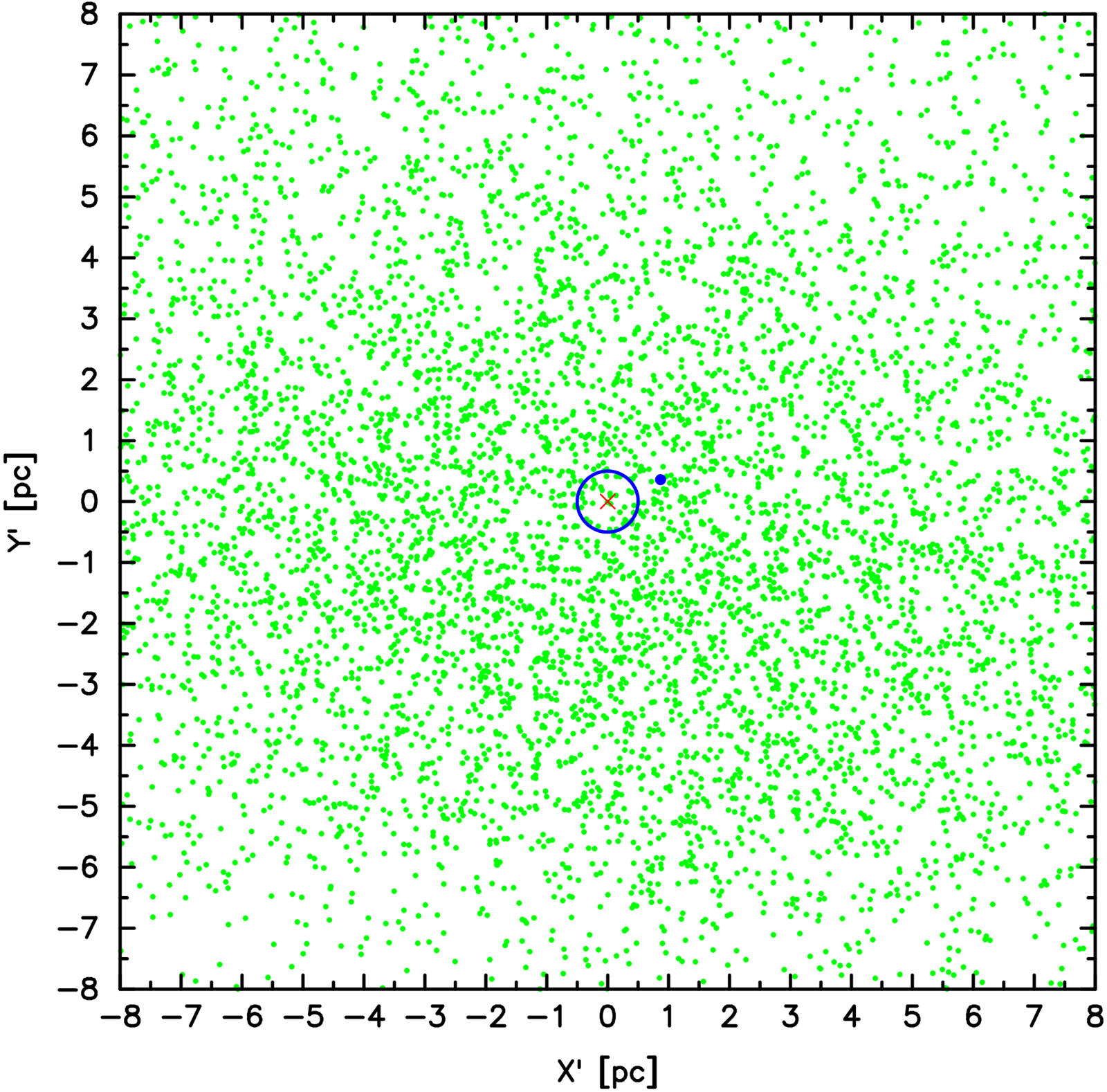}
        \caption{Spatial distribution of 10000 VS of P2004 at the moment of the closest approach in the plane of their maximum scatter.}Blue dot represents the nominal position of the star at the closest proximity and red cross represents the position of the Sun. The blue circle corresponds to a boundary of the Oort Cloud at 0.5 pc. Omitted: 3974 VSs that are too far from the Sun to be plotted.
        \label{P2004clones}
\end{figure}

\section{Stellar close flybys near the Sun based on \textit{Gaia} EDR3}
\label{final-passages}

For the StePPeD release 3.2 described in this paper, we prepared a significantly updated list of potential perturbers. First, based on our experience with cometary calculations \citep[see e.g. ][] {Dyb-Breiter:2021,dyb-kroli:2022}, we decided to temporary limit the list of stars and stellar systems to those that nominally pass the Sun closer than 2~pc and ignoring their uncertainties. It appeared that more distant perturbers practically do not disturb the LPC motion in a noticeable way. We systematically (at each StePPeD update) repeated a scanning of the calculations for the observed LPCs to check which stars perturb their motion in a noticeable manner. The last result of such a test can be found in \citet[][see their Table~2]{Dyb-Breiter:2021}. 

Moreover, due to  continuous high levels of uncertainties in some stellar data, in many cases, using a large list of potential perturbers seems inefficient and inappropriate at the moment. We also decided to temporary omit the stars that nominally pass close to the Sun but due to the current large distance from the Sun (or the low velocity) they must spend a long time to travel between their current position and the closest approach point, which strongly increases the propagation of their uncertainties.

Using the minimum distance threshold of 2~pc and temporarily excluding some promising stars on the basis of the subjective rank of the uncertainties (described above), we shortened the whole list to 155 objects. It consists of 142 stars (including seven binaries) from the previous list updated with the data from {\it Gaia} EDR3 and 11 new single stars and two binaries added on the basis of the new {\it Gaia} EDR3 data.

In Table \ref{tab-closest-past}, we present ten of the closest past encounters based on current stellar data.  In Table \ref{tab-closest-future}, we present ten of the closest future flybys. In the first two columns, we include our internal identifier and {\it Gaia} EDR3 number.  We describe the minimal Sun--star distance in two different ways.  The value presented in the third column, D$_{min}^{geom}$, is calculated in a special way: this is the distance from the Sun to the centroid of a cloud of 10\,000 VSs drawn according to the data uncertainties and using the respective covariance matrix.  As its uncertainty, we present a radius of a sphere around that centroid, which includes 68\% of VSs.  Uncertainties greater than the D$_{min}^{geom}$ directly indicate that the VSs cloud surrounds the Sun (see for example Table~\ref{tab-closest-past},  col.~3 for P0230 and its VSs cloud in Fig.~\ref{fig-four-stars}). In the fourth column, we present a formal statistical description of the Sun--star distance set for all VSs (D$_{min}^{stat}$) using the Gaussian mean and 1$\sigma$ uncertainty or the median and the upper and lower limits calculated on the basis of percentiles: 16\% and 84\% in the cases where the distribution is non-Gaussian. All these values are also expressed in parsecs. In exactly the same way, we present the epoch of the closest approach T$_{min}^{stat}$ (the fifth column, in Myr) and the relative velocity during the flyby V$_{rel}^{stat}$ (the sixth  column, in km\,s$^{-1}$. The last column in both tables presents the mass estimate.

In the case of P0403 in Table \ref{tab-closest-past},  we use as the initial conditions 
for the numerical integration the data from  \cite{Dupuy:2019} and due to the lack of the covariance matrix the drawing VSs technique was not possible. We just quoted here the uncertainties from this paper where available.

The comparison of the four representative examples of the VS clouds dispersion is presented in Fig.~\ref{fig-four-stars}. The first four stars from Table~\ref{tab-closest-past} are included in this figure.  The centroids of their corresponding clouds are shown, aligned along the horizontal line keeping the correct nominal distance from the Sun and maintaining the same scale of the VS clouds spread.  It should be stressed that the four stellar close approaches to the Sun presented in Fig.~\ref{fig-four-stars} took place in different epochs spread over a 2.5~Myr interval (see the  sixth column of Table~\ref{tab-closest-past}).
Without any doubt, the most important star in our new list is P0230 (HD~7977). It has moderately precise astrometric and kinematic data and RUWE=2.015. We discuss its very close flyby near the Sun in the next section.

In comparison with the StePPeD release 2.3, several important changes in the list of the closest stellar encounters to the Sun have appeared. What is probably the most important is the complete removal of the perturber P1079 (ALS~9243), discussed  in detail in WPD20. Its parallax changed from  10.5617$\pm$0.4027 to 0.16891$\pm$0.19025~mas, which moved this star to a large distance, additionally with a high uncertainty. The closest star-Sun approach of P0505 ( \object{Gaia~DR2 955098506408767360}) in the past is also cancelled due to the parallax change from 34.5051$\pm$0.6146 to 1.53497$\pm$0.02945~mas. These two stars were the most important perturbers in the StePPeD release 2.3 from the point of view of a dynamical evolution of LPCs, based on the {\it Gaia} DR2 results. 

\begin{table*}
        \caption{Ten past closest stellar passages near the Sun. Column descriptions are the same as in Table~\ref{tab-binaries-closest}. } \label{tab-closest-past}
        \begingroup
    \setlength{\tabcolsep}{6pt} 
    \renewcommand{\arraystretch}{1.7} 
        \begin{tabular}{c l c  r c c c r} 
                \hline 
                StePPeD & \multicolumn{1}{c}{{\it Gaia} EDR3}   & \multicolumn{1}{c}{geometry}         & \multicolumn{3}{c}{statistics}                        & mass estimate \\ 
                ID & \multicolumn{1}{c}{ID}                     & D$^{geom}_{min}$ [pc]                             & D$^{stat}_{min}$ [pc] & T$^{stat}_{min}$ [Myr] & V$^{stat}_{rel}$  [km\,s$^{-1}]$        & [M$_{\odot}$] \\
                \hline
                P0230 & 510911618569239040 & 0.014$\pm$0.040 & $0.032^{+0.027}_{-0.026}$ & $-2.471^{+0.026}_{-0.026}$& $30.65 \pm  0.29$ & 1.08 \\
                P0506 & 5571232118090082816 & 0.199$\pm$0.008 & 0.199 $\pm$ 0.006 & $-1.084 \pm 0.004$ & 90.16 $\pm$ 0.32 & 0.77 \\
                P0508 & 2946037094762449664 & 0.264$\pm$0.376 & $0.385^{+0.205}_{-0.177}$ & $-0.978^{+0.080}_{-0.094}$ & $39.95^{+1.89}_{-1.92}$ & 0.25 \\
                P0509 & 52952720512121856 & 0.318$\pm$0.153 & $0.334^{+0.099}_{-0.095}$ & $-0.670^{+0.050}_{-0.057}$ & $31.25 \pm 1.43$ & 1.46  \\
                P0403 & 3048443305671969152 & 0.335  & 0.333 $\pm$ 0.010&  $-0.081 \pm$ 0.001 & 82.5 & 0.16  \\
                P0533 & 3118526069444386944 & 0.504$\pm$0.155 & $0.510^{+0.120}_{-0.118}$ & $-3.179^{+0.085}_{-0.088}$ & $41.24    \pm 1.01 $ & 0.87 \\
        P0417 & 1281410781322153216 & 0.514$\pm$0.018 & 0.514 $\pm$ 0.013 & $-1.471 \pm 0.003$ & 32.7338 $\pm$ 0.0006 & 0.85  \\
                P0514 & 6608946489396474752 & 0.572$\pm$0.034 & 0.572 $\pm$ 0.027 & $-2.751^{+0.036}_{-0.036}$ & 46.00 $\pm$ 0.58 & 0.75  \\
                P0524 & 1949388868571283200 & 0.621$\pm$0.028 & $0.621^{+0.020}_{-0.018}$ & $-0.672^{+0.014}_{-0.015}$ & $355.2 \pm 7.0$ & 0.70  \\
                P0522 & 5261593808165974784 & 0.624$\pm$0.009 & $0.624^{+0.009}_{-0.008}$ & $-0.896^{+0.011}_{-0.011}$ & 72.71 $\pm$ 0.90 & 0.55 \\
                \hline
        \end{tabular}
        \endgroup
\end{table*}

\begin{table*}
        \caption{Ten future closest stellar passages near the Sun. Column descriptions are the same as in Table~\ref{tab-binaries-closest}. }\label{tab-closest-future}
        \begingroup
    \setlength{\tabcolsep}{6pt} 
    \renewcommand{\arraystretch}{1.7} 
        \begin{tabular}{c l c  r c c c r} 
                \hline 
                StePPeD & \multicolumn{1}{c}{{\it Gaia} EDR3}   & \multicolumn{1}{c}{geometry}         & \multicolumn{3}{c}{statistics}                        & mass estimate \\ 
                ID & \multicolumn{1}{c}{ID}                     & D$^{geom}_{min}$ [pc]                             & D$^{stat}_{min}$ [pc] & T$^{stat}_{min}$ [Myr] & V$^{stat}_{rel}$  [km\,s$^{-1}]$        & [M$_{\odot}$] \\
                \hline
                P0107 & 4270814637616488064 & 0.052$\pm$0.003 & 0.052 $\pm$ 0.002 & 1.290 $\pm$ 0.001 & 14.800 $\pm$ 0.006 & 0.65  \\
            P0551 & 1422721321394307456 & 0.173$\pm$0.313 & $0.279_{-0.136}^{+0.180}$ & $6.240_{-0.210}^{+0.228}$ & 39.27 $\pm$ 1.33 & 0.82  \\
            P2002 & 911145876981562496 & 0.405$\pm$0.266 & $0.448_{-0.151}^{+0.155}$ & $3.736_{-0.100}^{+0.103}$ & 31.27 $\pm$ 0.62  & 0.66 \\
            P0416 & 1952802469918554368 & 0.482$\pm$0.037 & $0.480_{-0.035}^{+0.040}$ & $0.070_{-0.005}^{+0.006}$ & $100.88_{-7.71}^{+7.90}$& 0.20  \\
            P0414 & 729885367894193280  & 0.511$\pm$0.401 & $0.301_{-0.163}^{+0.418}$ & $0.505_{-0.183}^{+0.570}$ & $95.69_{-50.45}^{+53.14}$& 0.08 \\
            P0504 & 4535062706661799168 & 0.548$\pm$5.923 & $4.375_{-2.295}^{+3.549}$ &$5.847_{-1.048}^{+1.494}$ & 29.69 $\pm$ 3.97 & 0.70 \\
            P0618 & 213090546082530816  &  0.672$\pm$0.262 & $0.711_{-0.147}^{+0.139}$ & $8.224_{-0.606}^{+0.730}$ & 23.75 $\pm$ 1.96 & 0.91  \\
            P0318 & 5469802896279029504 &  0.740$\pm$0.451 & $0.820_{-0.093}^{+0.179}$ & $13.763_{-1.141}^{+1.371}$ & $2.65_{-0.21}^{+0.21}$& 0.65  \\
            P0520 & 1722612190157591680 &  0.757$\pm$0.157 & $0.750_{-0.130}^{+0.150}$ & $4.485_{-0.172}^{+0.189}$ & 37.84 $\pm$ 1.50 & 0.60 \\
            P0567 & 4536673181955253504 &  0.763$\pm$0.226 & $0.772_{-0.143}^{+0.156}$ & $1.335_{-0.069}^{+0.074}$ & 73.70 $\pm$ 2.35 & 0.60  \\
                \hline
        \end{tabular}
        \endgroup
\end{table*}

For example, from among new stars, the closest nominal pass near the Sun  we obtained for P2001 (\object{Gaia EDR3 4312257326836128896}). It has a value of $D_{min}^{stat}=0.26$~pc, while the minimal distance from a linear approximation is $D_{min}^{lin}=53.12$~pc. Its present distance from the Sun is 166.28 pc. Despite its rather good astrometric data ($RUWE=0.80$), we cannot determine the flyby distance near the Sun reliably. The reason for that is the combination of the present distance of P2001 and its small radial velocity ($4.64$ $km\,s^{-1}$). The resulting journey time from the present position to the closest approach to the Sun becomes $45.86$ Myr. The uncertainties propagation on such a long interval of a numerical integration makes the final result highly uncertain. Figure~\ref{P2001clones} demonstrates the lack of reliable information on the close approach of this star. Green dots show the VSs at their closest position to the Sun. Due to the unreliable results, this star was removed from the final list.

On the other hand, we found several  promising objects, but they are characterised by a high level of inaccuracy on the part of their astrometry, which is reflected by RUWE parameter. For example, P2004 (\object{Gaia EDR3 3341225866116778880}) displays $D_{min}^{stat}=0.94$~pc and $T_{min}^{stat}=-9.06$~Myr. It has  $RUWE=13.22$ and its current distance is $373.33$~pc. As a result, the distribution of P2004 VSs shows that we cannot determine this star flyby near Sun in a reliable manner (see Fig.~\ref{P2004clones}).

\section{Important case of P0230}
\label{sect:P0230_case}

In the StePPeD release 2.3, based on {\it Gaia} DR2, we obtained a minimal distance between P0230 (HD 7977) and the Sun equal to 0.4~pc. The most important difference between DR2 and EDR3 astrometric data for this star appeared in the right ascension component of the the proper motion. In DR2, it was 0.559$\pm$0.040~mas yr$^{-1}$; however, in EDR3 it, becomes 0.144$\pm$0.024~mas yr$^{-1}$. The reason for such a substantial change could be attributed to the binarity of the star. However, we could not find any clues in this regard in any catalogue or other source. All uncertainties in EDR3 are smaller than those in DR2 and RUWE is only 2.015. Thus, it is clear that we should rely on EDR3 data for the moment.

\begin{figure}
        \includegraphics[angle=270, width=1.0 \columnwidth]{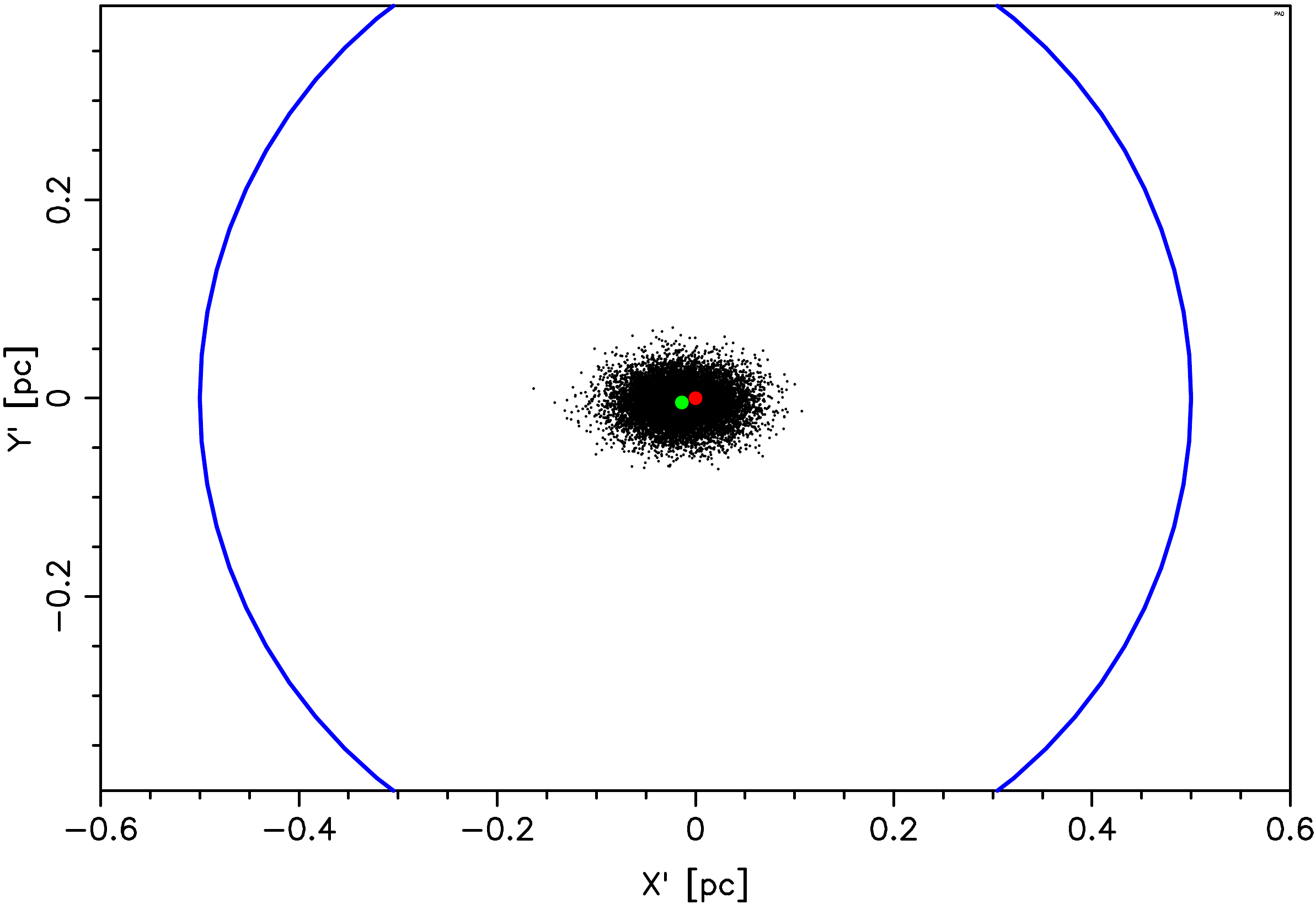}
        \caption{Closest pass of P0230 and its 10\,000 VSs near the Sun projected onto the VS maximum scatter plane. Red dot represents the Sun position, while the green one shows the nominal P0230 closest position. The blue circle depicts the 0.5~pc outer border of the Oort cometary cloud.The statistical description of the VS cloud can be found in Table~\ref{tab-closest-past}.}
        \label{fig-P0230clones}
\end{figure}

According to the new nominal data, P0230  passed the Sun 2.5 Myr ago at an extremely small distance of 0.014~pc ($\sim$3000~au). However, the uncertainties of its astrometric and kinematic data result in a considerable spread of its VS cloud that surrounds the Sun (as depicted in Fig.~\ref{fig-P0230clones}). 

There is a very important qualitative change resulting from the reduction of star-Sun nominal minimal distance from 0.4~pc to 0.014~pc. With the larger distance, passing near the outer border of the Oort cloud, this star can become an significant perturber only for some LPCs that appear near the stellar trajectory. Such a comet will be near its aphelion, so the orbit change might be spectacular. One such case of the comet C/2002~A3 has been described in detail in \cite{First-stars:2020}, based on {\it Gaia} DR2 data for P0230. 

When a star passes as close  as a few thousand astronomical units (or closer) to the Sun, the situation is different. In this case, the main perturbation is gained by the Sun and, as a result, its Galactic trajectory and its velocity are changed. This changes will be reflected in all heliocentric orbits of all Solar System bodies but on different levels. As we show in the next section, bodies in smaller orbits (e.g. planets) are practically not affected, but all LPCs on their highly elongated orbits would be strongly perturbed. The example cases of C/2014~UN$_{271}$ and C/2017~K2 were recently discussed by \cite{dyb-kroli:2022}.

Comparing this very close passage to that of the \object{Scholz's star} (P0403 in StePPeD), discovered by \citet{Mamajek:2015}, we can list several important differences. P0230 has passed ten times closer, with a much smaller relative velocity and it is considerably more massive. All these properties make P0230 appear a much more efficient comet perturber than the Scholz's star. Moreover, the passage of the Scholz's star appeared very recently (0.08~Myr ago) and if this star produced observable comets, we have to wait for them for thousands of years in the future. Conversely, P0230 passed 2.5 Myr ago (according to our results) and a lot of observed LPCs might have experienced a strong orbit change caused by this star  \citep[see e.g. ][]{dyb-kroli:2022}.

\section{Question of   rejecting the possibility of such a close stellar passage}

It is unlikely for stars to have passed too close to the Sun in the past, because they would have disturbed the planetary orbits. The closest acceptable star-Sun distance has been discussed in many papers. For example \cite{Morby-Levison:2004}, following \cite{Ida:2000},  discussed the passage of a solar mass star at a distance of 100 -- 200~au as the scenario of the origin of a strange orbit of Sedna. In another study, \cite{Adams:2006} estimated that a passage at a distance of 700-4000~au would not be disruptive for a planetary system. But in a review paper on cometary dynamics by \cite{Dones:2015}, these authors stated that a stellar passage at 400~au from the Sun would excite the Neptune eccentricity above 0.1, while a passage at 200~au from the Sun could have ejected Neptune from the Solar System. 

Taking the parameters of the P0230 nominal encounter with the Sun and the estimate of its mass, we decided to perform a series of simple tests to check how close to the Sun P0230 could have passed. We should keep in mind that due to the current P0230 data and their uncertainties the closest star-Sun distance could be arbitrarily small (see Fig.~\ref{fig-P0230clones}).

\begin{figure}
     \centering
        \includegraphics[width=\columnwidth]{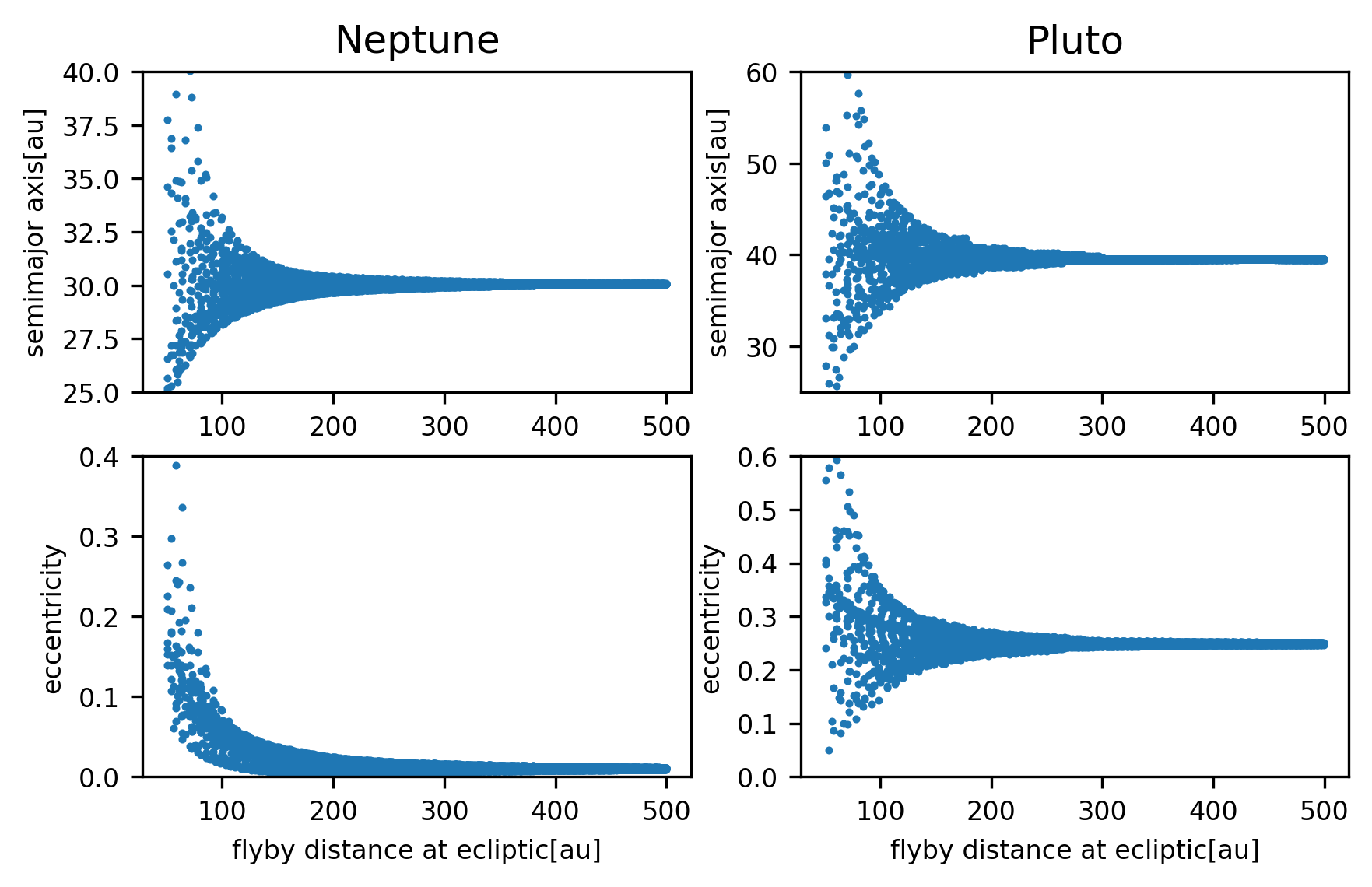}
        \caption{Effect of close flybys on orbits of Neptune and Pluto. Each point in the plots represents a single flyby of P0230 like star with different ecliptic plane crossing point. }
        \label{fig:Neptun_Pluto}
\end{figure}

\begin{figure}
     \centering
        \includegraphics[width=\columnwidth]{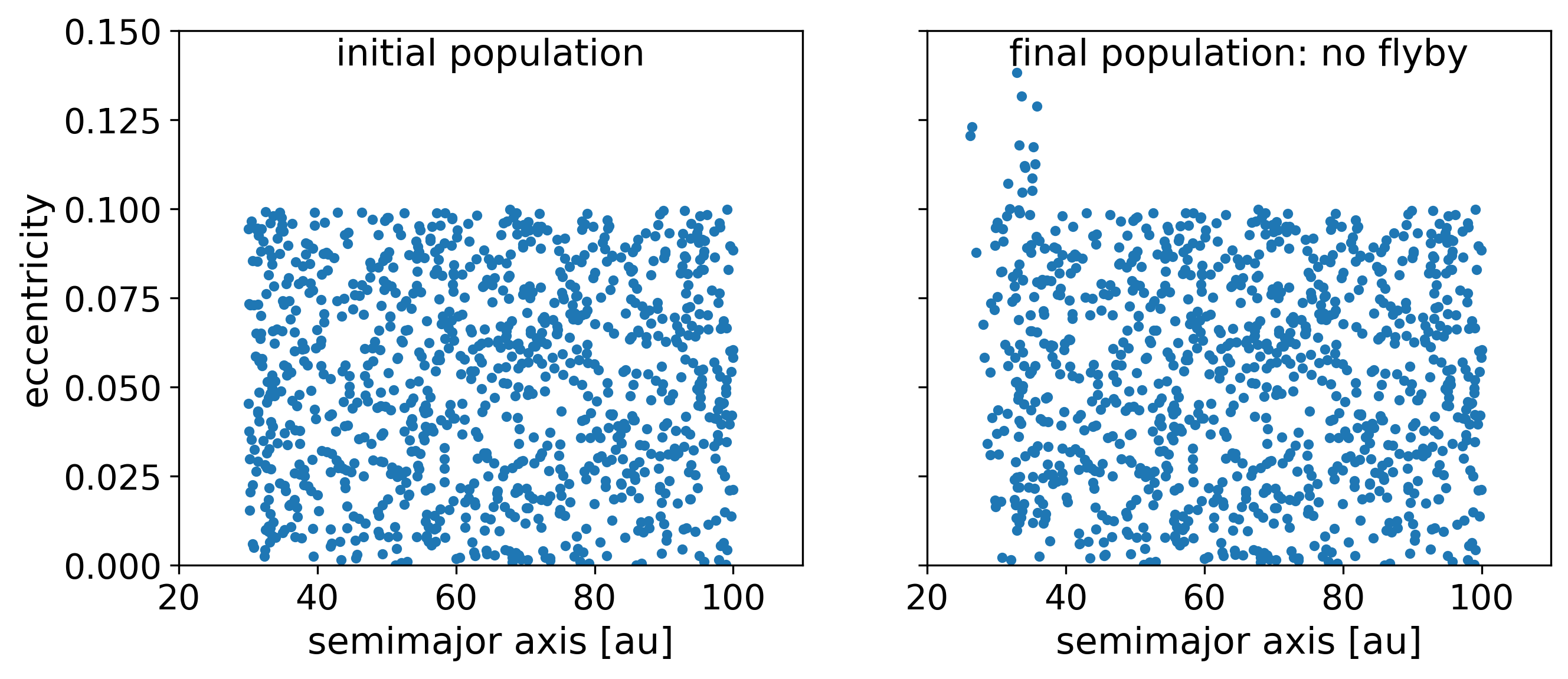}
        \includegraphics[width=\columnwidth]{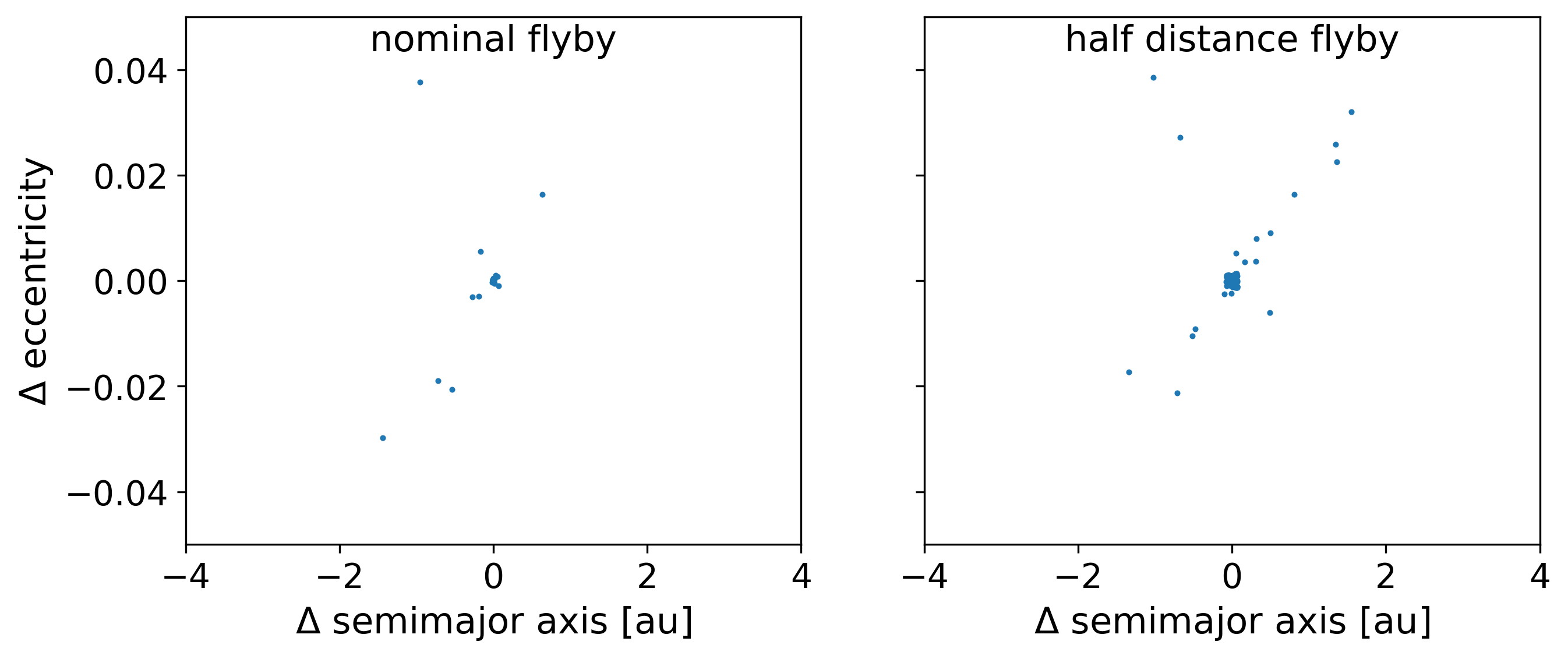}
        \includegraphics[width=\columnwidth]{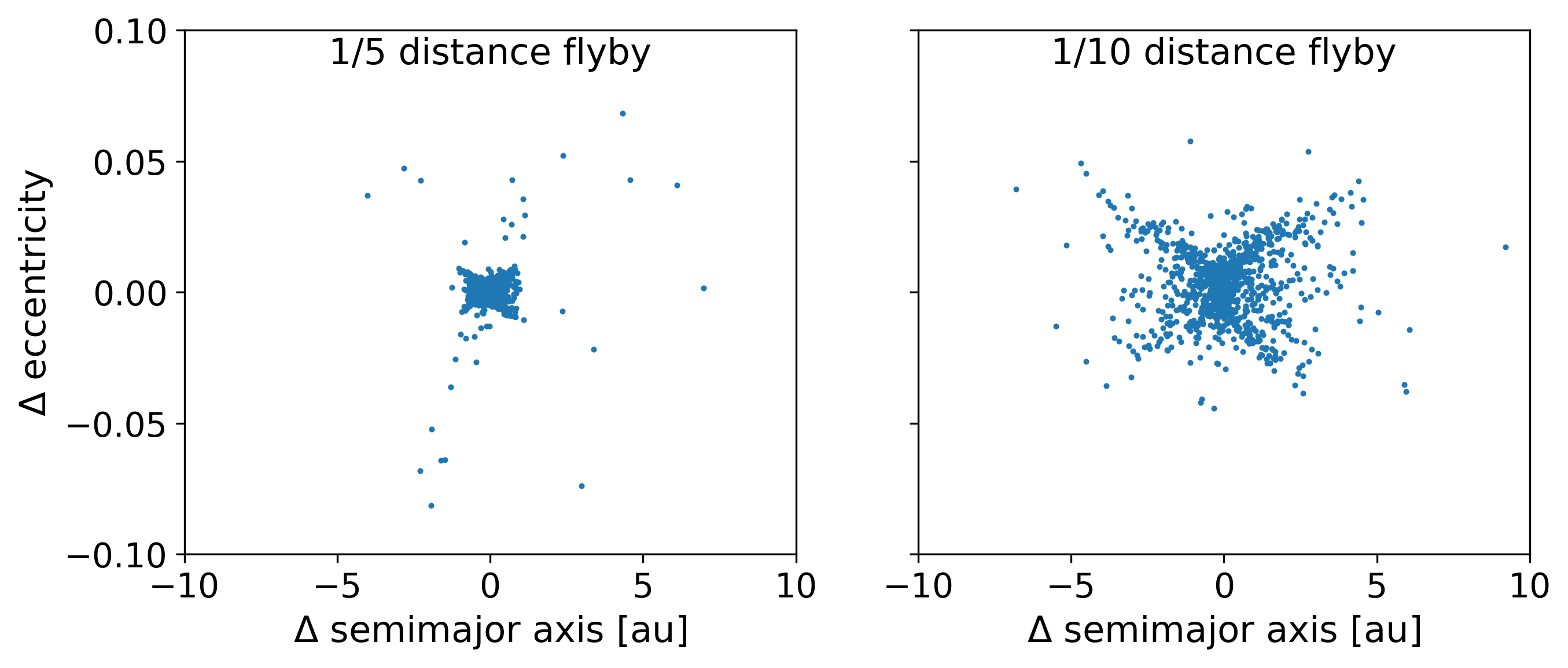}

        \caption{Effect of flybys on orbits of synthetic Kuiper Belt Object population. The initial orbits (top left) were simulated under gravitational effects of giant planets and Pluto without passing star (top right). Then we tested the same population adding flyby of P0230 at four different distances and plot differences in orbital elements caused by flyby compared to no flyby simulations (middle and bottom row, see the text for detailed descriptions).} \label{fig:KBOs}
\end{figure}

\begin{table}
        \caption{Effect of a passing star on the Solar System planets. The simulation spans 100\,000 years; a star passes its perihelion after 50\,000 years of simulation. In the consecutive columns we show: initial value of orbital elements and final values in three cases: with no passing star, with star passing at nominal distance (2\,940 au from Solar System barycenter) and with star passing ten times closer (294 au). }
        \label{tab-simulation-planets}
        \setlength{\tabcolsep}{4pt}
        \begin{tabular}{c c c c c c} 
                \hline \\
        object &  & initial & \multicolumn{3}{c}{after $10^5$ years}  \\
               &        &   & no star  & nominal & dist./10\\
        \hline
                &  a   &5.162     & 5.1774475     &5.1774485     & 5.1816781 \\
        Jupiter &  e  &0.044    &0.0386616    &0.0386610  & 0.0380938\\
                &  i    &1.30      &1.2510514      &1.2510599     & 1.2524185\\
        \hline
                &  a & 9.5258 & 9.5276163 & 9.5276170 & 9.5292422\\
        Saturn  &  e & 0.056 & 0.0771802 & 0.0771808 & 0.0773365\\
                &  i & 2.488 & 2.4780500 & 2.4781168 & 2.4849129\\
        \hline
                &  a & 19.1848 & 19.187124 & 19.187124 & 19.188364\\
        Uranus  &  e & 0.047 & 0.0313471 & 0.0313486 & 0.0321796\\
                &  i & 0.77& 1.3070023 & 1.3073480 & 1.3390967\\
        \hline
                &  a & 30.066 & 30.058894 & 30.058895 & 30.080055\\
        Neptune &  e & 0.0086 & 0.0117391  & 0.0117399 & 0.0128642\\
                &  i & 1.77& 1.8837079 & 1.8843004 & 1.9093493\\
        \hline
                & a &39.49 & 39.50569 & 39.50571 & 39.88493 \\
        Pluto   & e & 0.2490 & 0.2468883 & 0.2469112 & 0.2437073 \\
                & i & 17.14 & 17.260103  & 17.260642 & 17.272419 \\
        \hline
        \end{tabular}
\end{table}

To estimate the possible impact of the flyby as close as the nominal distance of P0230 on the Solar System, we used the Rebound N-body integrator software \citep{rebound:2012}. 
In our simulation we created a seven-body system with the Sun, four giant planets, Pluto, and a passing star. For initial positions and velocities we used a present-day Solar System JPL Horizon ephemeris and we added a passing star with the mass equal to the estimated P0230 mass (1.08 M$_\odot$). We choose a velocity value during the flyby that is equal to the value calculated for the nominal of P0230 and set the initial position of the star, so that it crossed the ecliptic plane halfway through the simulation duration. The first tests showed that the flyby at nominal distance has negligible effect on the giant planets and the impact on the Pluto orbit is also small (see Table~\ref{tab-simulation-planets}). The perturbation in orbital elements for Solar System objects created by star gravity is much smaller than the change in orbital elements caused by gravitational interactions between the giant planets. This results shows that such a close flyby is quite possible and would not affect the dynamics of planet in a significant way. With this result, we decided to check the impact of even closer stellar flybys to determine the minimum realistic distance at which this star could pass without disrupting our planetary system. We set the position of the star at the ecliptic plane at specific $(x,y,0)$ coordinates and moved it backward assuming a Keplerian hyperbolic orbit. We used the resulting position as the initial data in the simulation. We simulated the flybys with ecliptic crossing coordinates $x$ and $y$ ranging from $-500$ to $500$ au, creating the grid with step of $10$ au, excluding very close passages when a distance from the Solar System barycenter during ecliptic plane crossing is smaller than $50$ au. The obtained final eccentricity and semi-major axis of Neptune and Pluto are shown in Fig. \ref{fig:Neptun_Pluto}. These results show that even very close flybys are possible without any resulting disruption with regard to the orbits of the outer Solar System objects. We can assume that the safe minimum distance between the P0230 and the Solar System barycenter (measured when the star crosses the ecliptic plane) is about $300$~au -- ten times closer than the nominal value we obtained for P0230. At such a distance, the flyby causes the difference in the semi-major axis below 0.1~au when compared with the simulation without a passing star. The corresponding difference in eccentricity is smaller than 0.01. It is also worth noting that even a flyby at a smaller distance can be quite safe when occurring within the bounds of a favourable geometry.

During the simulation the Pluto was used as a simple indicator of the effect of P0230 on Kuiper Belt Objects (KBOs), but we decided to check more carefully the impact of such a stellar passage on the KBO population. To this aim, a randomised set of massless objects with a semi-major axis between 30 and 100 au on near circular orbits (eccentricities $e<0.1$) with small inclinations to the ecliptic plane ($i<10^\circ)$ generated using the uniform distribution of the orbital parameters was added to the simulation. This creates a synthetic disc representing the Kuiper Belt. As a point of reference the initial population (top left of Fig.~\ref{fig:KBOs}) was first simulated under gravitational effects of giant planets and Pluto and the resulting population is shown in top right of Fig.~\ref{fig:KBOs}. We note that some of the objects with a small semi-major axis increased their eccentricity even when the passing star was absent. This is the effect of orbital resonances or a close encounter with Pluto. 

After this initial test, we simulated the same synthetic population, but we added the P0230 star and plotted the difference in orbital elements between the simulations with and without flyby. 
We used nominal encounter scenario (middle left of Fig.~\ref{fig:KBOs}) and 
then we modified it multiplying the nominal minimal distance to 
the Sun by 0.5 (middle right of Fig.~\ref{fig:KBOs}), 0.2 (bottom left of Fig.~\ref{fig:KBOs}), or 0.1 (bottom right of Fig.~\ref{fig:KBOs}). The results shows that for majority of population the change in semi-major axis is smaller than 1~au and the change in eccentricity is smaller than 0.01 when the flyby distance is greater than 600~au. Even the closest approach we tested at about 300~au is not enough to affect the population in a significant way ($75.6$\% of the objects have a difference in the semi-major axis less than 2~au and less than 0.02 in eccentricity).

These simulations show that objects similar to P0230 could possibly fly near the Sun at a distance of about $3000$~au without creating a noticeable disturbance. Very close flybys (at a distance smaller than $500$~au) are also possible, however, the precise determination of their impact on Solar System will require more careful investigation considering the long-term Solar System stability and statistical properties of minor body population. Such an investigation is beyond the scope of this article.

\section{Other important potential perturbers in our list}
\label{sect:other_perturbers}

In addition to P0230 among the past close approaches to the Sun we should mention P0506, P0508, P0509, and P0417 as the stars that were recognised as the frequent perturbers of the observed LPCs past motion \citep{Dyb-Breiter:2021}. While for P0506 and P0417, we have quite precise stellar data, the uncertainties of P0509 are unsatisfactory large and of P0508 make its influence on comets practically undetermined \citep{dyb-kroli:2022}. We refer to Fig.~\ref{fig-four-stars} for more details as well. We should note here that both P0508 and P0509 are absent in {\it Gaia} EDR3, so we copied their astrometry from DR2.

Among the stars that will make the closest pass near the Sun in the future the record-holder is still P0504. This star is absent in {\it Gaia} EDR3 and we use DR2 astrometry. Its nominal smallest distance is 0.012~pc ($\sim$2400~au) but this result is highly unreliable. The cloud of its VSs is so great that that to enclose 68\% of them, we have to use a radius of almost 6~pc. The cloud is slightly asymmetric, which causes its centroid to be 0.55~pc from the Sun and the close passage parameters strongly dependent on the applied model of a Galactic potential. Detailed statistics of this approach can be found in the first row of Table~\ref{tab-closest-future}. 

In the second row of this table one can found parameters of the close passage of the well known star Gliese~710 (P0107 in StePPeD). This star was recognised many years ago as approaching closely to the Sun in the future, see for example \cite{mullari-o:1996, dyb-kan:1999, garcia-sanchez:2001}. The parameters of its predicted close passage were updated based on the {\it Gaia} DR1 catalogue in \cite{berski-dyb:2016}. In the StePPeD version 2.3, based on {\it Gaia} DR2, they obtained a passage at 0.054~pc in 1.29~Myr. In the present paper the updated values are almost identical, only the minimum distance decreased to 0.052~pc. This star has the smallest uncertainties among the studied stars so precise calculations of its future influence on cometary orbits can be performed \citep[see e.g.][]{dyb-kroli:2022}. 

In our current list, there is also another very promising candidate for a significant comet motion perturber, namely, P0551. According to the available data, it will pass near the Sun nominally at a distance below 0.2~pc in 6 Myr, however, the uncertainties of its astrometry are far too large to treat this result as definitive. It would be also valuable to obtain more precise radial velocity for another candidate, P0414; however, its small mass makes it rather marginally important from a point of view of perturbing comet motion. Another interesting case is P0318. It will pas the Sun in over 13 Myr but a very good astrometry and precise radial velocity allowed us to obtain rather reliable parameters of its future approach to the Sun.

\section{Summary, conclusions, and prospects}

In this paper, we describe several aspects of the update of the StePPeD database of the potential comet motion perturbers accounting for the {\it Gaia} EDR3 data. The first step was to update parameters of the close approaches to the Sun for all stars included in the previous StePPeD releases. We use the minimal star-Sun distance as a filtering tool when searching for potential stellar perturbers. It appeared that significant changes in astrometric data for a large percentage of these stars resulted in the exclusion of over one third of the total from further consideration.

In this situation, it was necessary to search the whole {\it Gaia} EDR3 catalogue for new candidates. Basing on our previous experience, we decided to repeat the whole procedure described in WDP20 but with one important change: using the linear approximation as the first filter, we allow for a much greater star-Sun distance than the 10~pc threshold used in some previous papers \citep{dyb-berski:2015,Bailer-Jones:2018}. This allows us to find a lot of potential new candidates. After the second filtering, by means of the numerical integration of each star with the Sun under the Galactic potential, we obtained over 500 stars passing closer than 5~pc from the Sun and more than half of them have a minimal distance obtained from a linear approximation larger than 10~pc.

Next, we closely examined each star from this shorter list searching for probable binary or multiple system. To this aim, we mainly used data from \cite{million-binaries} but we also manually searched for secondaries directly in the {\it Gaia} EDR3 catalogue. More than 10\% of the stars appeared to be members of multiple systems, which makes it necessary to calculate their centre of mass motion and check again for their proximity to the Sun.

Using the very rich material presented in \cite{million-binaries}, we also decided for the first time to search for binaries overlooked in all earlier papers on this subject because none of their components, treated as a single star, did not appear in the list of stars approaching the Sun.
During this first attempt, we isolated over 15\,000 systems for which the radial velocity of both components is known. Then we filtered this list by checking whether it is possible to obtain a close passage near the Sun for some mass ratio between the components. This left only 156 candidates on our list. For these, we searched for available mass estimates and we analysed in detail the possibility of the close approach by varying the component masses within their confidence intervals. This complicated and time-consuming procedure resulted in only two new potential perturbers added to the final list.

We also studied how our results might depend on the chosen model of the Galactic potential. Our test with five other models show that for the majority of studied objects the differences are below 10\%.

When preparing the final list of potential stellar perturbers of the LPCs motion we decided to apply the new and more restricted threshold of 2~pc as the maximal allowed distance from the Sun at the moment of the closest approach. This decision resulted from several cometary orbital calculations showing that more distant passages are rather irrelevant,   \citep[see e.g.][]{Dyb-Breiter:2021,dyb-kroli:2022}. Using this new threshold, we compiled stars from the previous list, new single stars, and new multiple systems and we obtained a list of 155 potential perturbers for the new release 3.2 of the StePPeD database.

In this new release, apart from the large amount of information previously presented, we contribute two important new results: for all objects, we calculated the uncertainty of the parameters of the close approach to the Sun and we include a graphical presentation of this uncertainty by means of the picture of 10\,000 VSs of the star in question drawn according from the appropriate covariance matrix and stopped at their closest approach to the Sun. All the presented results are based on the {\it Gaia} EDR3 data (where available) and all other information (e.g. mass estimates) are also updated to accommodate the latest published values.

In the new list of potential perturbers of the LPCs motion, we found several interesting objects. The most important seems to be the star HD~7977 (P0230 in the StePPeD database) which nominally passed the Sun 2.5~Myr ago at an extremely small distance of 0.014~pc ($\sim$3000~au). We performed several simple tests checking the effect of such a close star passage on the outer planetary system bodies. It seems that the outer Solar System planets are save even during much closer passages but LPCs were all strongly perturbed. A similar situation will occur in the future during the passage of Gliese~710 (P0107).

Concerning the completeness of the StePPeD database, we know that there are missing stars in Gaia EDR3, both faint red and highly luminous. 
For a number of years, we monitored the Simbad database and Vizier catalogues for stars with known 
parallax and radial velocity and checked all new objects found. Generally we are not particularly interested in the smallest objects (e.g. brown dwarfs) because their small masses make them very inefficient perturbers of cometary motion. Instead we carefully check all massive stars absent in the Gaia catalogue, for example, we carefully 
studied the 'missing' list of stars presented with the Gaia Catalogue of Nearby Stars \citep{gcns}. 

We are fully aware that a new release of data from the {\it Gaia} mission is expected in a few months. We hope to see a great improvement in the astrometry precision and a large number of the new radial velocity measurement. The expected large amount of new data will require a considerable time to obtain an updated list of potential perturbers. With this in mind, we decided to prepare and publish the StePPeD update based on the currently available data. All the methods developed for this purpose and tested practically in WDP20 and the current paper might be used in the next update iteration.

\begin{acknowledgements}

We would like to express our thanks to the anonymous referee for so many so detailed and so helpful suggestions and comments. These allowed us to greatly improve this paper. The calculations which led to this work were partially supported by the project  "GAVIP-GC: processing resources for Gaia data analysis" funded by European Space Agency (4000120180/17/NL/CBi).

\end{acknowledgements}

\bibliographystyle{aa} 
\bibliography{PAD31} 

\end{document}